\documentclass{elsarticle}

\usepackage{mathptmx}
\usepackage{soul}\setuldepth{article}

\def\hb{\hbox to 11.5 cm{}}

\usepackage{soul}
\usepackage{url}
\usepackage[hidelinks]{hyperref}
\usepackage[utf8]{inputenc}
\usepackage[small]{caption}
\usepackage{graphicx}
\usepackage{amsmath,amssymb}
\usepackage{booktabs}
\usepackage{placeins}
\usepackage{bbm}
\usepackage{caption}
\usepackage{subcaption}
\usepackage{comment}
\usepackage{makecell, multirow}
\usepackage{array}
\usepackage{placeins}

\usepackage{color}
\definecolor{Orange}{rgb}{1,0.5,0}
\definecolor{Red}{rgb}{1,0,0}
\definecolor{Blue}{rgb}{0,0,1}

\newcommand\ChangeRT[1]{\noalign{\hrule height #1}}

\journal{PAIS 2022}

\begin{document}

\begin{frontmatter}



\title{Automating the resolution of flight conflicts: Deep reinforcement learning in service of air traffic controllers}


\author[inst1]{George Vouros}
\author[inst1]{George Papadopoulos}
\author[inst1]{Alevizos Bastas}
\affiliation[inst1]{organization={University of Piraeus}, 
addressline={Piraeus}, 
            country={Greece}
            }

\author[inst2]{Jose Manuel Cordero}
\author[inst2]{Rub{\'e}n Rodrigez Rodrigez}
\affiliation[inst2]{organization={CRIDA}, addressline={Madrid}, 
            country={Spain}
            }

\begin{abstract}
Dense and complex air traffic scenarios require  higher levels of automation than those exhibited by tactical conflict detection and resolution (CD\&R) tools that  air traffic controllers (ATCO) use today. However, the air traffic control (ATC) domain, being safety critical, requires  AI systems to which operators are comfortable to relinquishing control, guaranteeing operational integrity and automation adoption.  Two major factors towards this goal are quality of solutions, and transparency in decision making. This paper proposes using a graph convolutional reinforcement learning  method operating in a multiagent setting where each agent (flight) performs a CD\&R task, jointly with other agents. We show that this method can provide high-quality solutions with respect to stakeholders interests (air traffic controllers and airspace users), addressing operational transparency issues. 
\end{abstract}




\end{frontmatter}

\section{Introduction}
Aiming to contribute to the automation of operations in real-life, complex, safety-critical  settings, AI systems need to meet domain-specific objectives: Systems must be effective to make accurate predictions and  prescribe actions  that resolve problematic situations w.r.t the interests of stakeholders; e.g., without increasing cost of operations or compromising safety. In addition, they need to meet objectives regarding human performance and engagement: Human operators should be comfortable relinquishing control to a system \cite{mythos}, and should be able to take control safely at any time. Transparency of decision making is crucial here, especially when system's responses do not comply with operators' usual practices or intuition. Operators need to associate system responses to the real-life situations and inspect the consequences of alternatives suggested. This is important towards improving trust, safety and accountability.

Air Traffic Management (ATM) is the integrated management of air traffic and airspace safely, economically and efficiently, through the provision of infrastructure facilities and seamless services, in collaboration with all parties and involving airborne and ground-based functions (ICAO Doc. 4444).  Air Traffic Control (ATC) in this domain, according to ICAO Annex 11, is a service provided for the purpose of: a) preventing collisions: 1) between aircraft, and 2) on the maneuvering area between aircraft and obstructions; and b) expediting and maintaining an orderly flow of air traffic. These tasks are related to airspace capacities declared by the appropriate authority. The traffic volume must not exceed capacities, which should be utilized to the maximum extent. Indeed, the airspace capacity is a crucial factor for the efficiency and safety of operations, defined by the ability of humans to control airspace volume (i.e. an airspace sector, called Area of controllers' responsibility (AoR)). Increase of the airspace capacity, thus density and complexity of traffic, without compromising safety and efficiency of flights, introduce challenging issues in the aviation industry, where  AI can provide solutions.  

Our aim is to advance the automation for  conflict detection and resolution (CD\&R) among flights, contributing to operational integrity and automation adoption.  To achieve this challenging goal, we propose building a solution based on a graph convolutional reinforcement learning (RL)  method capable of operating in real-world multiagent settings where agents (flights) cooperate, but without explicit communication. We show that the proposed method can provide high-quality solutions, addressing  ATCO transparency requirements.
The contributions that this work makes are as follows:

- It proposes an enhanced graph convolutional reinforcement learning  method operating in a multiagent setting, where each agent (flight) performs a CD\&R task, jointly with other agents;
 
- It evaluates the proposed method in real-world scenarios, providing evidence for the quality of solutions w.r.t. the interests of ATCO and airspace users (AU, i.e. airlines);
 
- It addresses issues of transparency in decision making, according to operational requirements and constraints.

The paper is structured as follows: Section 2 provides preliminary knowledge for the CD\&R task and describes proposals towards automating it, using RL methods. Section 3 specifies the CD\&R task as a multiagent learning problem, while section 4 describes in detail the CD\&R method we propose. Section 5 presents how transparency requirements are met, while Section 6 provides evidence on the quality of solutions provided, w.r.t. the interests of ATCOs and AUs. 

\section{Preliminaries and related work}
To maintain the risk of collision between aircraft in acceptable levels, the ATM system requires that the aircraft does not breach certain separation minima both at the horizontal and vertical axes. The minimum prescribed horizontal separation when using surveillance systems is 5 nautical miles (5NM)  (ICAO Doc 4444). This may be further reduced or increased under specific conditions. The specified minimum vertical separation for Instrument Flight Rules (IFR) flights is 1000 ft (300 m) below FL290 and 2000 ft (600 m) from FL290 and above. When Reduced Vertical Separation Minima (RVSM) apply, this changes to 1000 ft (300 m) below FL410 and 2000 ft (600 m) from FL410 and above.\footnote{https://www.skybrary.aero/index.php/Separation\_Standards} A \textit{loss of separation} is defined as the violation of separation minima in controlled airspaces, whereas a \textit{conflict} is defined as a {\em predicted} violation of the separation minima. \textit{Alerts} are conflicts estimated to occur within a restricted time horizon (e.g. within 10 seconds).
 
 Nowadays conflicts are detected and resolved by a Planner Controller (PC) and an Executive Controller (EC), which is an ATCO, in a per sector (volumes segregating the airspace) basis, in two respective phases: the planning and the tactical phase, respectively. 
 While conflict detection and resolution in the planning phase may suggest changes in the flight plan (i.e. the flight intended trajectory, as filed even before take-off and updated until landing), in the tactical phase it mainly implies changes of the actual flight trajectory, given the trajectory flown up to the current time point, the last flight plan, and/or prediction(s) on the evolution of the trajectory from that time point and on. Prediction is crucial and implies uncertainties in trajectory evolution: In this work, focusing on the tactical phase of operations,  trajectories are projected into a limited future time horizon in a nominal way, as existing operational tools do.

 ECs (ATCO) detect and resolve conflicts in their AoRs,  also coordinating with the ATCO of the downstream sectors: Coordination mostly concerns the AoR exit point conditions, ensuring safe entry of flights in the downstream sector. While safety is the top priority, ATCO, close to the interests of AUs, should also aim to increase the efficiency of the flights, without adding extra nautical miles, flight time, or increasing fuel consumption and emissions due to abrupt changes in speed,  flight level, or other reasons.
 
 RL has been already proposed, among other methods, for automating  the CD\&R task. However, various efforts present specific limitations and/or make crucial assumptions about agents, thus suggesting solutions with a delta from real-world settings. Transparency has not been addressed adequately, yet.

Authors in \cite{pham2019machine} model the single-agent problem
(the ownship flight). The agent in the presence of uncertainty, chooses the lateral maneuver
that resolves a conflict with another flight, assuming adherence of all other flights to
their planned route. The model
was trained by using Deep Q-Network (DQN) and Deep Deterministic
Policy Gradient (DDPG). Although the results show the potential of the proposal, this work, assuming a single-agent decision process,  does not consider agents' cooperation, which is important to address safety and flight efficiency.
The Multi-agent Reinforcement Learning (MARL) approach in \cite{brittain2019autonomous} follows a parameter-sharing approach: The model consists
of a distributed actor-critic neural network, trained with  the Proximal
Policy Optimisation (PPO) algorithm. Agents however consider only  speed adjustments (i.e. a limited repertoire of actions) to maintain safe separation
while moving along the planned 2D route.    Assuming the route identifier as input, implies that the policy model is not transferable to agents executing different routes in various airspace. Authors in \cite{li2019optimizing} propose a model-based approach assuming perfect knowledge of environment dynamics. Again, agents perform resolutions in the lateral plane. For conflicts with several agents, the problem
is split into pair-wise conflict resolution sub-problems.
Very close to our approach, offering also some inspiration to our problem formulation and use of edges among conflicting agents, authors in \cite{Dalmau2020ATC} present a recommendation tool based on MARL to support ATCO
in complex traffic scenarios. The model consists
of a distributed Message Passing Actor Critic Model exploiting Message Passing Neural Networks \cite{10.5555/3305381.3305512}. Parameters
are shared among all agents and are learned using the PPO algorithm. This method allows flights (agents) to exchange information through
a communication protocol before proposing a joint action that
promotes flight efficiency and penalises dangerous situations. The policy function is
trained in a controlled simulation
environment, while limited transparency is provided.
In \cite{ghosh2020deep}, authors propose a method that combines Kernel Based Stochastic Factorization and a deep MARL method using the PPO algorithm. These methods are combined by another deep policy model that at each timestep decides which of the two models to choose. The proposed method considers flights in the en-route phase only, considering only the possibility of speed change. However, the method reports good scalability, tested in a simulated setting.
Focusing on explainability, authors in \cite{EgorovATM21} propose an approach based on a lattice-space exploration process. Actions are represented by 3D tuples changing the course, the speed, and the altitude of flights. The action space is modeled as a lattice that can be pruned and explored in a breadth-first bottom-up manner, giving priority to actions that impose small changes in the course,  speed and altitude, creating small deviations from the flight plan. This lattice-based approach inherently provides explanations for choosing an action, in contrast to any other action, but it is not clear whether explanations provided address ATCO transparency requirements adequately.

Here, we advance previous efforts in several dimensions: We propose using a deep MARL  method that inherently supports agents' cooperation, DGN \cite{jiang2020graph}, enhances with features describing conflicts among agents, modeled as edges. Agents do share policy model parameters, and they do take advantage of observations of their neighbors, jointly with own observations and edges. Furthermore, agents are capable to detect and resolve conflicts in 3D with a rich repertoire of actions. Finally, our approach has been designed with transparency in mind, so as to satisfy requirements stated by ATCO, and has been trained and tested in real-world scenarios.

\section{Problem Specification}
The CD\&R task involves a number of flights in an AoR. The task is about  detecting at any time point $t$ the conflicts that may occur between flights, and decide whether and  what resolution actions should be applied to the conflicting flights. 

Casting this problem into a multiagent problem, we consider that each agent $i \in A$ represents one of the $N$ flights in the AoR, or in any downstream sector, for any of the flights crossing the AoR. We consider the set of \textit{Relevant AoRs (RAoRs)} as the set of  potential downstream sectors and the AoR.

Given the trajectory $T_i$ of agent $i$ within the RAoRs, we define the set of \textit{neighboring agents} to be the set of \textit{conflicting flights to $T_i$ in RAoRs at a specific time point t}. I.e., the set of agents that will potentially be in loss of separation with $i$, as assessed at time point $t$. These are denoted $Neigh(i,AoR,t) \subseteq A$.

Agent $i$ has to react and resolve all conflicts with $Neigh(i,AoR,t)$, deciding whether it will apply any  resolution action  at at $t$, and what this action should be.
Specifically, as considered here, the following actions comprise the repertoire of agents' actions:
(a) \textit{Flight Level change}, where the agent changes its current flight level one level up /down, assuming a vertical speed $17$/$-17$ feet/s for ascending/descending course;
(b) \textit{Course change}, where the available changes of agent's course are $10, -10, 20, -20$ degrees;
(c) \textit{Horizontal speed change}, where the available changes of agent's horizontal speed are $-3.6008$ or $3.6008$ m/s, for deceleration or acceleration, respectively;
(d) \textit{Direct to waypoint}, where the agent can choose one of the next four flight plan waypoints; and 
(e) \textit{No action}, where the agent continues its current course without any change. 

Actually, an ATCO determines a resolution action and its duration.  If the conflict persists after the execution of the action, the ATCO issues further instructions, but if the situation worsens while executing the action, an intervention is possible. Here, actions duration is added as an option, expanding further the action space. 
According to the domain experts, the accepted range of actions' duration is 1-3 minutes. Therefore, agents can choose the duration of each action among four values: 30, 60, 120 and 180 seconds. It must be noted that duration is not decided for ``direct to waypoint" and ``flight level change" types of  actions, since these actions are executed within the time span required by the aircraft\footnote{This work does not consider different aircraft abilities and limitations. Adding such features in the state and conditioning actions under specific aircraft abilities, would refine solutions, without affecting much the design of the proposed method.} to reach the target state.  This results into a repertoire of 32 discrete actions: An agent at each timestep has to decide on the specific action (e.g. the specific change on speed) and its duration, where it applies.

The problem is formulated as a Decentralized Partially Observable Markov Decision Process (Dec-POMDP), where at each timestep $t$ each agent $i$ receives a local observation $o^t_i$, takes an action $a^t_i$, and gets an individual reward $r^t_i$. The objective is to maximize the sum of all agents expected returns.

A local observation of an agent is a vector comprising the following features:

- $Nalt = alt/max_{alt}$, where $alt$ is the agent's current altitude in feet and $max_{alt}$ is a normalization factor,

- $\cos \chi$ and $\sin \chi$, where $\chi$ is the bearing of the aircraft, i.e. the angle of the agent's course w.r.t North, in degrees,

- $Nh_{speed}=\frac{h_{speed}-min_{h_{speed}}}{max_{h_{speed}}-min_{h_{speed}}}$, where $h_{speed}$ is the magnitude of the agent's horizontal speed in m/s, $max_{h_{speed}}$ and $min_{h_{speed}}$ are precalculated factors,

- $\cos(\chi-\psi)$ and $\sin(\chi-\psi)$, where $\psi$ is the relative bearing of the agent w.r.t. the intended AoR exit point, according to the flight plan,

- $NdistExitPoint= d_{exit}/D_{exit}$, where $d_{exit}$ is the horizontal distance of the agent w.r.t. the AoR exit point in meters, and $D_{exit}$ is a normalization factor,

- $NaltDiffExitPoint=\frac{\lvert alt - alt_{Exit Point} \rvert}{max_{\lvert alt - alt_{Exit Point} \rvert}}$, where $\lvert alt - alt_{Exit Point} \rvert$ is the absolute difference in feet between the agent's  altitude at the exit point and the filed (according to the flight plan) altitude at the exit point,

- $\cos(dcourse_{wp})$ and $\sin(dcourse_{wp})$ for each one of the next four waypoints $wp=1,2,3...$, where $dcourse_{wp}$ is the angle of the current agent's course w.r.t. the course that the agent must follow to reach the corresponding waypoint,

- $NdistWaypoint= h_{d_{wp}}/\mathit{HD}$ for each one of the next four waypoints, where $h_{d_{wp}}$ is the horizontal distance in meters between the current agent's position and the position of the corresponding waypoint. $\mathit{HD}$ is a normalization factor,

- $NaltDiffWaypoint=v_{d_{wp}}/\mathit{VD}$ for each one of the next four waypoints, where $v_{d_{wp}}$ is the vertical distance in feet between the  agent's altitude at the waypoint and the filed (according to the flight plan) altitude at  that  waypoint. $\mathit{VD}$ is a normalization factor. 

\begin{figure}
\centerline{\includegraphics[width=18pc]{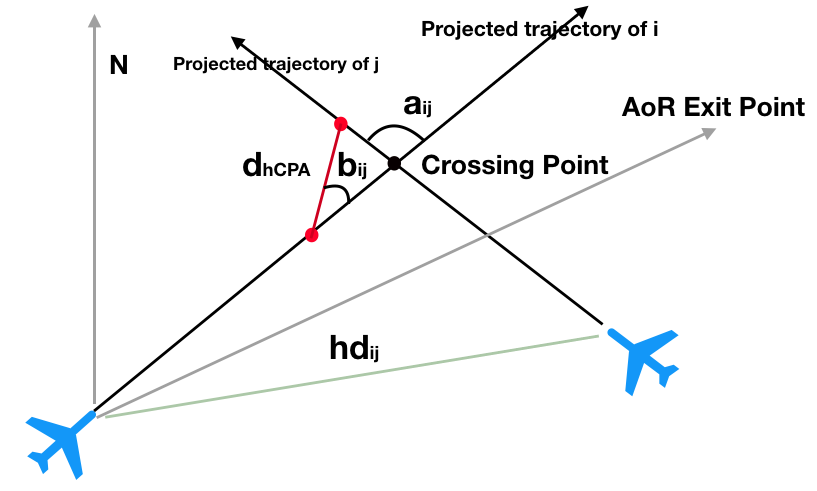}}
\caption{CPA geometry.}
\label{fig:CPA}
\end{figure}

In addition to these observations, each agent $i$ maintains a vector $e_{ij}$  with any agent $j \in Neigh(i,AoR,t)$, comprising $ij$ edge features that mostly depend on the geometry at the Closest Point of
Approach (CPA), i.e., at the point where agent $i$ will be (or have been) 
closer to $j$  (shown in Figure \ref{fig:CPA} with a red dot in the trajectory of $i$ and discussed in Section 4.1). These edge features are the following:  

- $Nt_{CPA}=t_{\mathit{CPA}}/\mathit{T_{CPA}}$, where $t_{\mathit{CPA}}$ is the time required in seconds for agent $i$ to reach the $\mathit{CPA}$ with agent $j$ and $\mathit{T_{CPA}}$ is a normalization factor,

- $Nd_{h_{CPA}}=d_{h_{\mathit{CPA}}}/{D_{h_{\mathit{CPA}}}}$, where $d_{h_\mathit{CPA}}$ is the horizontal distance in meters between agents $i$ and $j$ at the $\mathit{CPA}$ and $D_{h_\mathit{CPA}}$ is a normalization factor, 

- $\cos a_{i,j}$ and $\sin a_{i,j}$, where $a_{i,j}$ is the intersection angle in degrees between $i$ and $j$,

- $\cos b_{i,j}$ and $\sin b_{i,j}$, where $b_{i,j}$ is the relative bearing of  $i$ w.r.t. to  $j$ at the $\mathit{CPA}$,

- $Nd_{v_{CPA}}=v_{d_{\mathit{CPA}}}/V_{d_{\mathit{CPA}}}$, where $v_{d_{\mathit{CPA}}}$ is the vertical distance in feet between  $i$ and $j$ at the the $\mathit{CPA}$, and $V_{d_{\mathit{CPA}}}$ is a normalization factor,

- $Nd_{cp}=d_{cp}/D_{cp}$, where $d_{cp}$ is the distance in meters between  $i$ and $j$ when any of them passes the crossing point first, and $D_{cp}$ is a normalization factor,

- $Nt_{cp}=t_{cp}/T_{cp}$, where $t_{cp}$ is the time required in seconds for any of  $i$ and $j$ to pass the crossing point first, and $T_{cp}$ is a normalization factor, 

- $Nd_h(i,j)=h_{d_{i,j}}/\mathit{HD}$, where $h_{d_{i,j}}$ is the current horizontal distance in meters between agents $i$ and $j$,

- $Nd_v(i,j)=v_{d_{i,j}}/\mathit{VD}$, where $v_{d_{i,j}}$ is the vertical distance in feet between $i$ and $j$.

\section{Automating CD\&R}

\begin{figure}
\centerline{\includegraphics[width=18pc]{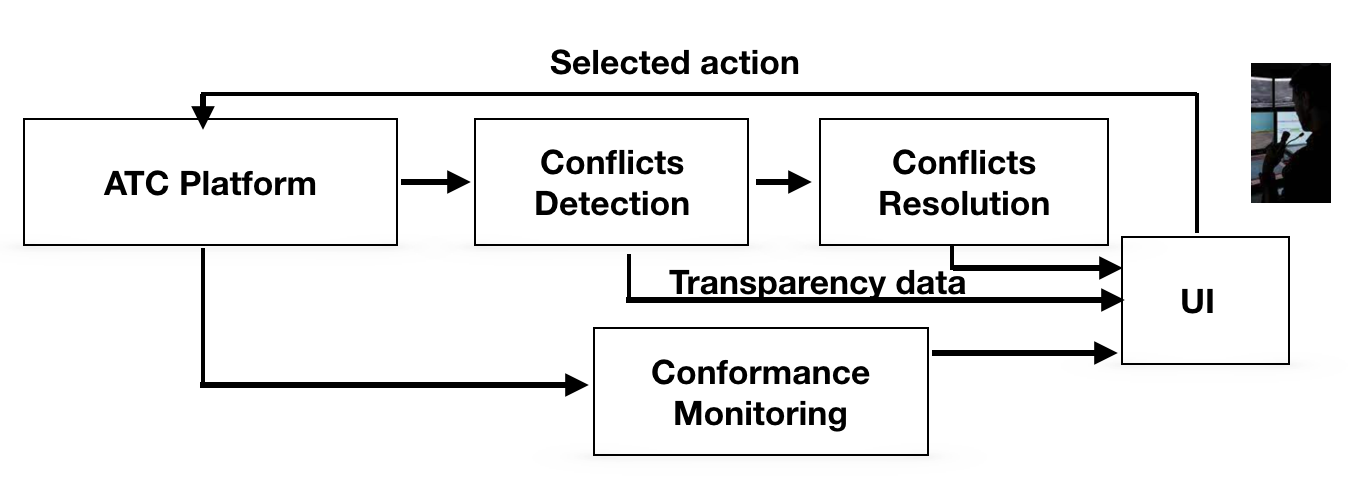}}
\caption{CD\&R System Overall Architecture.}
\label{fig:arch}
\end{figure}

The overall system implemented for automating CD\&R comprises three main components (Figure \ref{fig:arch}): The conflicts detection, the conformance monitoring, and the conflicts resolution component. In addition to these, there is a component that provides basic interaction capabilities with the human operator, integrating and presenting transparency data, in parallel to the information provided by the operational platform. 
The system is integrated with the Spanish ATCO operational platform SACTA and it is fed with a stream of data from that platform, providing every 30 seconds updates of flights' radar tracks and updates of flight plans.

Minimum interaction facilities provided allow the  system to operate either as an advisor (in that case the ATCO has to chose an action to be applied), or in full automation mode (the system takes the initiative to act). In any case, 
the ATCO gets appropriate information on the rationale for resolution actions, and he/she may unfold the transparency data, to get  further information on the situation and on the consequences of actions, in cases where strict operational constraints allow.

The conformance monitoring component, monitors whether an aircraft conforms with conflict resolution actions prescribed, in a short time horizon of 30 seconds. 

Since the focus is on conflicts detection and resolution, subsequent sections describe the corresponding components in detail.

\subsection{Detecting conflicts}
\begin{figure}
\centerline{\includegraphics[width=18pc]{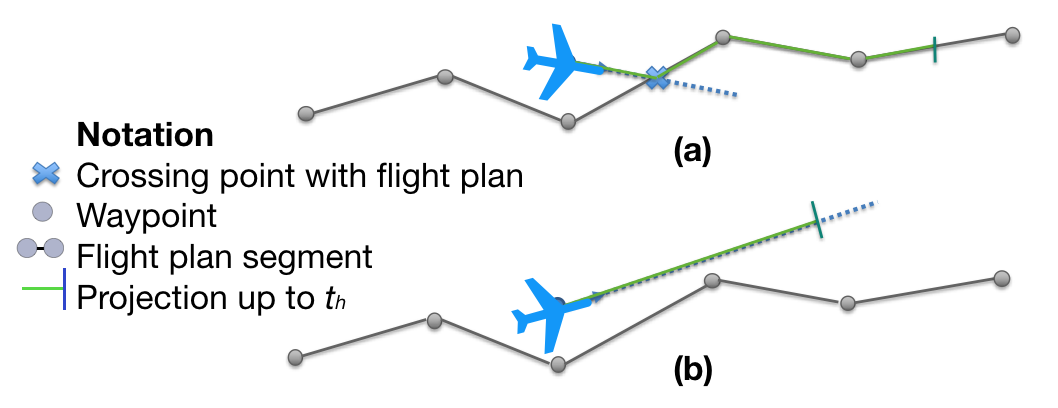}}
\caption{Cases for projecting flights' trajectories up to $t_h$ minutes.}
\label{fig:projection}
\end{figure}

To detect conflicts, given the state of a flight at a time point $t$, its trajectory is projected into the future for a specific time horizon $t_h$ following a nominal approach. As shown  in Figure \ref{fig:projection}, if it is assessed that the flight follows its flight plan, the projection is estimated according to the flight plan, else, it is estimated according to the actual flight's course. 

Given the line segment defined by projecting the aircraft trajectory for $t_h$ minutes, we consider that the flight follows its flight plan if either (a) the shortest distance between the aircraft's 2D position and the flight plan horizontal profile is less than a distance threshold $d_h$, and the difference between the aircraft and the flight plan's course - at the closest point to the aircraft's position- is less than $c_h$; or (b) the line segment defining the projection, intersects with the flight plan's horizontal profile. In any other case we consider that the flight deviates from its flight plan. In both cases we assume that the flight will retain its current vertical and horizontal speed for $t_h$. 

Having decided on the projection of the aircraft trajectory,  the detection of conflicts considers the following two cases: (a) Both flights have zero vertical speed, (b) one of the flights has a vertical speed greater than zero. In case (a), if  the vertical separation minimum 
between aircraft is violated, the flights'  projection segments 
that intersect in the temporal dimension are detected, and the horizontal Closest Point of Approach (CPA) among he flights is computed using the methodology presented in \cite{pham2019machine}. 
In case (b) the CPA in the vertical axis (vCPA) is computed. In case the vertical separation minimum is violated at vCPA, the method checks near the vCPA to find any point at which the horizontal separation minimum is violated. If the vertical and the horizontal separation minimum are violated at a point,  a conflict is detected.

The values of the thresholds after study of conflicts' conditions and consultation with experts are as follows: 
The time horizon $t_h$ is equal to the time needed to reach the next flight level, if vertical speed is $>0$. Otherwise, it is equal to 10min. The distance threshold $d_h$ is set to 2km  and the course difference $c_h$ is set to 20 degrees.

\subsection{Resolving conflicts using graph convolutional reinforcement learning.}
\label{subsection: CR}

Cooperation among agents is crucial, as safety - being the top priority of all stakeholders in ATM - and flight efficiency - being the main interest of AU - demands it in an inherently multi-agent setting. According to the graph convolutional reinforcement learning method DGN \cite{jiang2020graph}, agents (flights) are  able to learn cooperative policies by focusing on their neighbors (i.e. those with whom they do interact) instead of taking into account all the existing agents in their environment. DGN uses a deep Q network and is trained end-to-end. DGN shares weights among all agents  - making it easy to scale, abstracts the mutual interplay between agents by relation kernels, and extracts latent features by convolution. The multi-agent interaction setting is modelled as a graph that DGN considers dynamic: This is of crucial importance to the CD\&R task, as  neighboring agents change. 

DGN comprises three main modules: An encoder for encoding agent's $i$ observations $o^t_i$ at any time step $t$, projecting them in a higher dimensional space, two convolutional layers for implicit communication  among the neighboring agents, expanding their receptive fields, and the Q network which takes as input all features from the encoded observations and the outputs of the preceding convolutional layers \cite{jiang2020graph}.

For the representation of neighboring flights, the adjacency matrix $C^t_i$ of  agent $i$ is of size $(|Neigh(i,AoR,t)|+1)\times N$. Each row represents either $i$ itself, or $j \in Neigh(i,AoR,t)$. The neighbor agents are sorted in the following order: First are those that  are in loss of separation with $i$. These are further ranked by their Euclidean distance to $i$. These are followed by the agents causing an alert with $i$ and are further ordered based on the time  they need to reach the CPA, $t_{\mathit{CPA}}$. Agents that are in conflict with $i$ follow, and these are ordered according to $t_{\mathit{CPA}}$. Finally, the neighbors which are in conflict but with a negative $t_{\mathit{CPA}}$  have the lowest priority.   
By merging all $C^t_i$ at $t$, we get the matrix $C^t$ of size $N \times (|Neigh(i,AoR,t)|+1)\times N$. 

\begin{figure}
\centerline{\includegraphics[width=30pc]{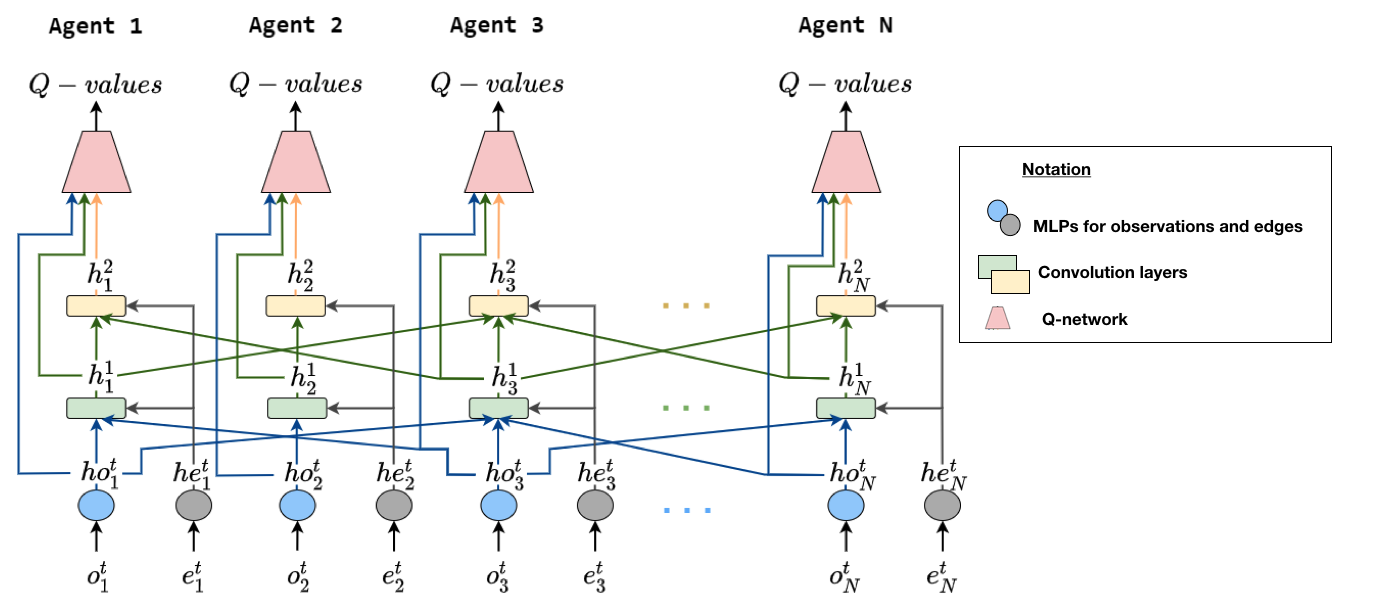}}
\caption{DGN model extended with edges. Connections show agents' interaction, according to their adjacency matrix, at time step $\mathit{t}$.} 
\label{fig:dgn_model}
\end{figure}

Having extended DGN to consider  vectors $e^t_{ij}$ comprising features of an edge between neighbor (conflicting) agents  $i$ and $j$, in addition to observations of $i$ at each time step $t$, the elements of $e^t_{ij}$ are sorted as elements in $C^t_i$ do. The first element of $e^t_{ij}$ consists of predefined values for agent $i$, in order to facilitate calculations. 
Edge vectors are encoded by a distinct MLP.   
The encoded vectors of observations and edges ($ho^t_j$ and $he^t_{ij}$, respectively) are concatenated into vectors in $hc^t_{ij}$, which are stacked horizontally, obtaining a matrix of such vectors per agent $i$. These last operations (concatenation and stacking) are performed by the first convolution layer, as illustrated in Figure \ref{fig:dgn_model}. Each vector $he^t_i$ shown in Figure \ref{fig:dgn_model} contains all $he^t_{ij}, \  \forall \  j \in Neigh(i,AoR,t) \cup \{i\}$.  

The next steps of the algorithm comprise the computations in the two convolution layers. The convolution kernel is the multi-head dot-product attention kernel, which can be computed for each agent $i$ and its neighbors as follows:
\small
$$a^l_{ij} = \frac{exp\Big(\big(W^l_{ii} \cdot (K^l_{ij})^\intercal\big) \times (1/\sqrt{d_{K^l_{ij}}})\Big)}{\sum_{j} exp\Big(\big(W^l_{ii} \cdot (K^l_{ij})^\intercal\big) \times (1/\sqrt{d_{K^l_{ij}}})\Big)}
$$
\normalsize
where, $j$ varies in $\big(Neigh(i,AoR,t) \cup \{i\}\big)$, $l$ denotes the corresponding convolution layer, $d_{K^l_{ij}}$ is equal to the size of each head in $K^l_{ij}$, 
$W^l_{ii} = f^l_W(hc^t_{ii})$
and
$K^l_{ij} = f^l_K(hc^t_{ij})$.
Here,  $f^l_W$ and $f^l_K$ are one-layer MLPs. It should be clarified that when $l>1$, the vector $hc^t_{ij}$ is obtained by concatenating the output of the preceding convolution layer, $h^{l-1}_j$, along with the corresponding encoded edges $he^t_{ij}$. $W^l_{ii}$ is a matrix of size $m \times d_{W^l_{ij}}$, with $d_{W^l_{ij}}=d_{K^l_{ij}}$ and $m$ the number of attention heads. The same holds for  $K^l_{ij}$. The attention values $a^l_{ij}$ show the significance that the neighbor-agent $j$ has for agent $i$ decision, as reflected in the outcome of the convolution layer, denoted by $h^l_i$.
The final step, as shown in Figure \ref{fig:dgn_model}, 
concatenates the outputs of the first and second convolution layer, $h^1_i$ and $h^2_i$ respectively, as well as the encoded observations $ho^t_i$, and feeds the result into the $Q$ network.

During training, at each timestep $t$ a prioritized experience replay buffer stores tuples of the form  $(O, E,Ac,\mathcal{O}, \mathcal{E},R,C)$ ($t$ is not denoted for simplicity), where $O = \{o_1,\dots ,o_N\}$ is the set of agents' observations and $E$ is the set of edges vectors, $Ac = \{a_1,\dots ,a_N\}$ is the set of agents actions, $\mathcal{O'}= \{o'_1,\dots ,o'_N\}$ is the set of observations and $\mathcal{E'}$ is the set of edges vectors resulting after the joint execution of actions, $R = \{r_1,\dots,r_N\}$ is the set of rewards received by agents, and $C = \{C_1,\dots ,C_N\}$ is the set of adjacency matrix per agent. 

Sampling  a  minibatch of size $S$ from the replay buffer, the  loss minimized is 
\small

    $\mathcal{L}(\theta)$=$\frac{1}{S}\sum_S\frac{1}{N}\sum^N_{i=1}\big((y_i-Q(O_{i,c},a_i;\theta)\big)^2$
,
\normalsize
where, $y_i = r_i + \gamma max_a Q(O_{i,C},a;\theta')$, $O_{i,C} \subseteq O$ denotes the set of observations of the agents in $i$ receptive fields determined by $C$, $\gamma$ is the discount factor. DGN uses an online $Q$ network parameterized by $\theta$,  and a target $Q$ network parameterized by $\theta$'. Both approximators take $O_{i,C}$ as input and output a $Q$ value per action $a_i$. The gradients of $Q$-loss are accumulated to update the $\theta$ parameters of the online $Q$ approximator. The target network parameters are updated by $\theta'= \beta\theta' +(1 - \beta)\theta'$, where $\beta$ is a hyperparameter. 

Any action executed can change the graph representing neighbor flights at a next timestep. As suggested in \cite{jiang2020graph} $C$ is kept unchanged in two successive timesteps when computing the $Q$-loss in training, to ease  learning.  

The reward function for each agent $i$, given the state resulting after applying any of the actions in a state, is an additive function that depends on the following terms: (a) the caused change in shift from the exit point, as provided by $\chi-\psi$, $NdistExitPoint$, and $NaltDiffExitPoint$, (b) the change in $\chi$, (c) the occurrence of speed change (horizontal and vertical), and (d) the number of losses of separation and alerts caused. 

Specifically, the reward function given state \textit{s'} resulting after applying action \textit{a} in state \textit{s} is as follows:
\small
\begin{align*}
    R(s,a,s')=&-1*\frac{\Delta\chi}{F}-0.5*\frac{|\chi-\psi|}{\pi}-0.5*NaltDiffExitPoint \\
    &-1*distExitPoint-1*\mathbbm{1}_{\Delta_{hspeed}}!=0-1*\frac{\Delta_{vspeed}}{V}\\
    &-10*NumberOfAlerts-5*NumberOfLosses
\end{align*}
\normalsize

\noindent where $\mathbbm{1}_{\Delta_{hspeed}}!=0$ indicates whether there is any change in the horizontal speed, and $V$ and $F$ are normalization factors.

\section{Transparency in decision making}

Data provision for transparency is triggered by conflicts. Transparency requirements gathered from ATCO specify that for any specific conflict, they do need to be provided with (a) data regarding the conflict detected and how it has been assessed, (b) data regarding resolution action per aircraft and major factors driving  decisions w.r.t. the interests of ATCO and AU, as well as (c) deviations from the flight plan and violation of AoR exit point constraints.  This supports them to assess a  situation and decide on the appropriate action: Specifically, they do need information that reflects the system's view of the conflicts, of the foreseen effects of actions, as these are estimated by the system, and a  ranking of the alternative actions per conflict instance. The effects of a resolution action concern foreseen conflicts caused by the resolution action, nautical miles added to the flight trajectory and additional flight time, deviation from the planned course, foreseen alerts or losses of separation, due to traffic. These should be provided with respect to operational desiderata (specified below), and at the appropriate level of detail. Specifically, given that there is little or no time for ATCO to exploit and explore detailed explanations while monitoring and resolving safety critical situations, transparency must satisfy Grice's maxims: Be informative and give the  quantity of information needed and no more, provide adequate information that describes adequately the situation assessed by the system, be relevant, clear and  brief. However, the decision to ask for further information about the proposals/decisions of the system should remain with human operators: If detailed transparency data is provided all the time, the increase in the level of workload of the operators might not compensate the benefit provided by the transparency itself.

Regarding any \textit{conflict detected at time point } $t$, the following information is provided for situation awareness: \textbf{(a)} The discrepancy between the actual track of each of the aircraft and of the corresponding flight plan in 3D (horizontal and vertical dimensions), at the first point of their intersection, or at the closest point between the aircraft course and the flight plan after $t$, as determined by applying the process specified in Section 4.1; \textbf{(b)} Whether the conflict has been detected using a projection of the actual trajectory or using the flight plan; \textbf{(c)} The features regarding the geometry of the CPA detected; \textbf{(d)} Foreseen horizontal, vertical and temporal distance to the CPA  for each of the aircraft; \textbf{(e)} The first and last points of conflict detected (these are not used by the conflict resolution method, as it uses the distance and the time to the CPA); \textbf{(f)} The horizontal and vertical distance of aircraft at the current position; and \textbf{(g)} Whether each aircraft is in the climb or descend phase.

For the \textit{resolution of conflicts}, besides the decided resolution action and its duration,  the system provides the following information: \textbf{(a)}  Nautical miles that may be added to the trajectory due to any change caused by an action; \textbf{(b)} The deviation from the trajectory course caused by the action; \textbf{(c)} Additional conflicts that may be caused by the resolution action: For each of these conflicts, conflict-relevant information is provided; \textbf{(d)}
 Losses of separations with other flights, or alerts foreseen, due to the resolution action; \textbf{(e)} In cases of multiple conflicts involving a specific agent, the system provides an attention hitmap, showing the attention the agent  to the neighboring flights, prioritizing them for the resolution of conflicts; and \textbf{(f)}
 the deviation  from  the  AoR  exit  point  in 3D.

\section{Evaluation setting and results}

\begin{table}[]
    \centering
    \footnotesize
    \begin{tabular}{c|c}
       Hyperparameter  & DGN \\ \hline
        discount ($\gamma$) & 0.96 \\
        batch size & 256\\
        buffer capacity & $2 \times 10^5$ \\
        $\beta$ & 0.01 \\
        $\epsilon$, min $\epsilon$ and decay & 0.6/0.001/0.996 \\
        number of neighbors & 3\\
        number of convolutional layers & 2\\
        m (number of attention heads) & 8\\
        $d_{K^l_{ij}}$ & 16\\
        number of encoder MLP layers & 2 \\
        number of encoder MLP units  (DGN/DGN+SE) & (512,128)/(512,256) \\
        encoder MLP activation & ReLU \\
        $Q$ network & affine transformation \\
        initializer & random normal \\
        train steps per episode & 80 \\
        exploration episodes (SeqN/AllN) & 3000/6000 \\
        exploitation episodes (SeqN/AllN) & 1000/2000 \\ 
        episodes before training & 200 \\ \hline
        PER starting $\beta$ & 0.4 \\
        PER max $\beta$ & 1 \\
        PER $\beta$ decay rate per step & 0.0025 \\
        PER $\epsilon$ & 0.05\\ 
        PER $\alpha$ & 0.6 \\ \hline
    \end{tabular}
    \normalsize
    \caption{Hyperparameters (PER: Prioritized Experience Replay)}
    \label{tab:hyperparams}
\end{table}

This section provides results on real-world scenarios where AoR is a sector near the Barcelona airport. Scenarios are constructed by choosing a flight crossing the AoR and a downstream sector (let us call it ``main flight") in a time span of $Duration_{main}$ seconds. The ``main flight" is chosen so as to present at least one conflict. Flights added in the scenario are all flights that cross the relevant AoRs (RAoR) in a time interval that temporally intersects with the time span of the main flight. Data about each of the scenarios is provided in Table \ref{tab:scenarios} in article's supplementary material. Scenarios involve varying number of flights,  thus various number of initial conflicts, losses and alerts (resolution actions may cause the appearance of new events) and have varying duration. The scenario ID is constructed by a timestamp and the AoR ID and it is of the form timestamp-AoRiD. Table \ref{tab:scenarios} provides the number of flights, as well as the initial number of conflicts, initial alerts and losses per scenario, and the duration of the scenario.

The proposed enhanced DGN method (DGN enhanced with edges) is compared against the original DGN method that exploits the information encoded in edges in a rather direct way. 
The naive but straightforward manner  is to include edges' $\{e^t_{ij}| j \in (Neigh(i,AoR,t)\}$ features in agent $i$ observations. This variation uses  one MLP encoder in contrast to the enhanced method where two distinct MLP encoders (one for the observations and one for the edges) are used.  
However, we have carefully designed the encoder (two-layer MLP with 512 and 256 neurons, respectively) of this variation, to get almost the same number of parameters as in the case of the two encoders, and the length of the output vector to be equal to the length of the vector comprising the output of the two encoders after their concatenation. 

This method variation is denoted as DGN+SE, and the policy models learned by this method have the ``+SE" indication. The comparison between methods aims to show the importance of edges as considered here, providing to each agent information regarding the CPA geometry with any neighbor agents. 
The conjecture is that  this naive way to incorporate edges' features in agents' observations does not allow the attention mechanism to work properly, because the combined encoded values do not refer exclusively to agents' $i$ and $j$. Hence, the agents will not be able to construct advanced collaborative strategies as they will be incapable of paying the proper attention to each of their neighbors. Technical details on DGN+SE are included in the supplementary material.

Models have been trained in different scenarios following one of two alternative training patterns. First we have trained a model providing samples from a set of N scenarios: The resulting model is denoted AllN. As an alternative,  we have constructed a model trained using samples from M batches of N scenarios, in sequence, as follows: First, we train a model (named 1SeqN) using a first batch of N scenarios. 1SeqN has been re-trained in a second batch of scenarios resulting in model 2SeqN, and so on, until the final model MSeqN. Thus, MSeqN has been trained in M$\times$N scenarios. Models All(M$\times$N) and MSeqN have been trained in the same scenarios. However, MSeqN models have been trained using a much larger number of episodes compared to All(M$\times$N). To alleviate the effect of this, each model is trained on a batch of N scenarios with 4000 episodes (3000 for exploration and 1000 for exploitation), in contrast to All(M$\times$N) which is trained with 8000 episodes (6000 for exploration and 2000 for exploitation). In any of the training patterns, the scenarios are considered in the order that appear in Table \ref{tab:scenarios}. Specifically, Table \ref{tab:scenarios} shows information about 42 scenarios. All30 and 5Seq6 have been trained using the first 30 of those scenarios, while they have been tested on the last 6. Similarly, 6Seq6 has been trained using the first 36 scenarios, and it has been tested in the remaining 6.

The DGN hyperparameters values set for the experiments are shown in Table \ref{tab:hyperparams}. 

Results from  policy models All30, 5Seq6 and 6Seq6 for DGN with edges are shown in Table \ref{tab:results}. 
Specifically,  Table \ref{tab:results} shows the percentage of conflicts resolved in training and testing the different models, the average number of resolution actions applied per scenario, and the average 
NM added to trajectories affected by resolution actions. It must be noted that 5Seq6 is superior to All30 regarding the percentage of conflicts resolved in the testing scenarios, showing the benefits of using the MSeqN training pattern. This superiority is further enhanced, by re-training the 5Seq6 model using the next batch of 6 scenarios. Indeed, 6Seq6 manages to solve 90\% of conflicts, compared to 66.67\% of 5Seq6. In addition to this, the efficacy of MSeq6 models concerning the percentage of conflicts resolved increases consistently as M increases. 
 
\begin{figure}
     \centering
     \begin{subfigure}[b]{0.40\textwidth}
         \centering
         \includegraphics[width=15pc]{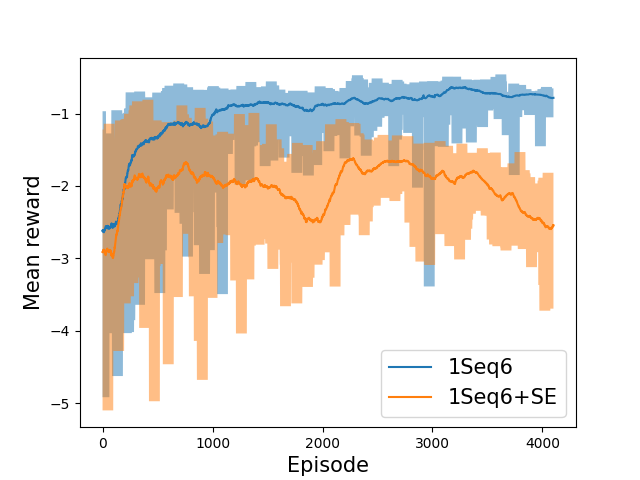}
     \end{subfigure}
     \hfill
     \begin{subfigure}[b]{0.40\textwidth}
         \centering
         \includegraphics[width=15pc]{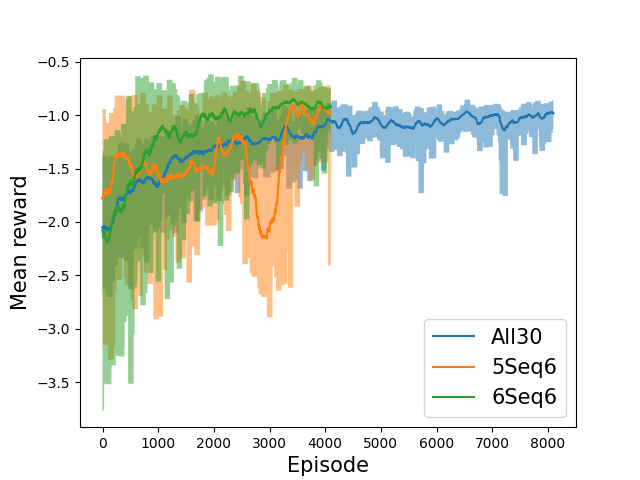}
     \end{subfigure} \\
     \begin{subfigure}[b]{0.40\textwidth}
         \centering
         \includegraphics[width=15pc]{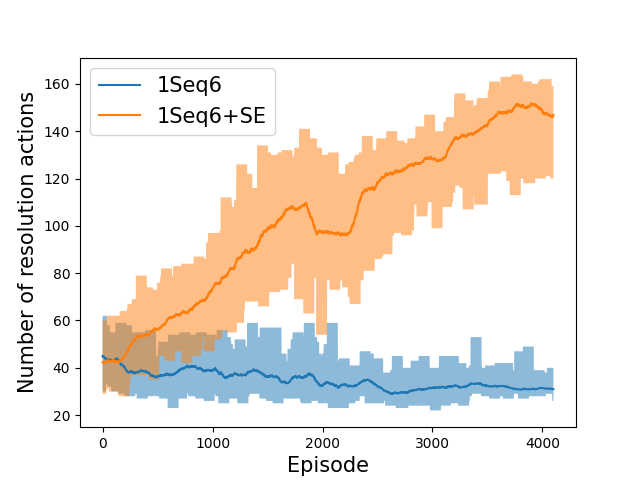}
     \end{subfigure}
     \hfill
     \begin{subfigure}[b]{0.40\textwidth}
         \centering
         \includegraphics[width=15pc]{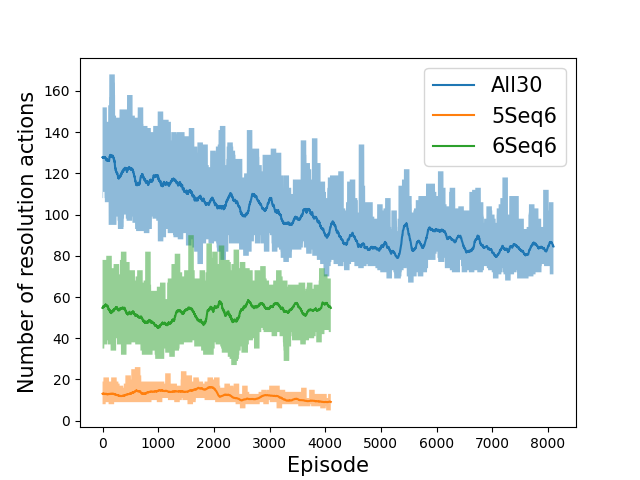}
     \end{subfigure} \\
      \begin{subfigure}[b]{0.40\textwidth}
         \centering
         \includegraphics[width=15pc]{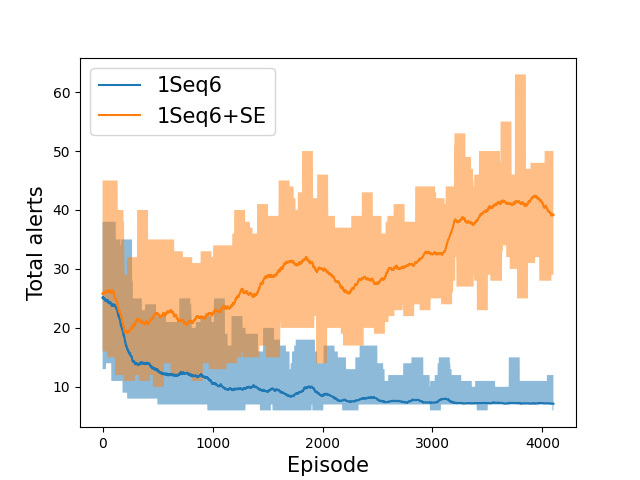}
     \end{subfigure}
     \hfill
     \begin{subfigure}[b]{0.40\textwidth}
         \centering
         \includegraphics[width=15pc]{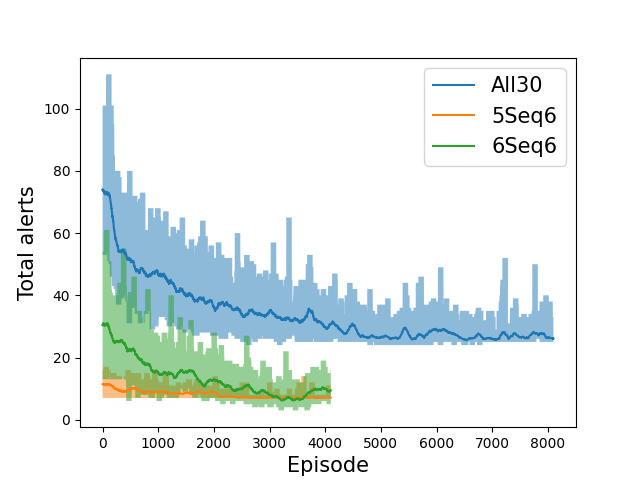}
     \end{subfigure} \\
     \begin{subfigure}[b]{0.40\textwidth}
         \centering
         \includegraphics[width=15pc]{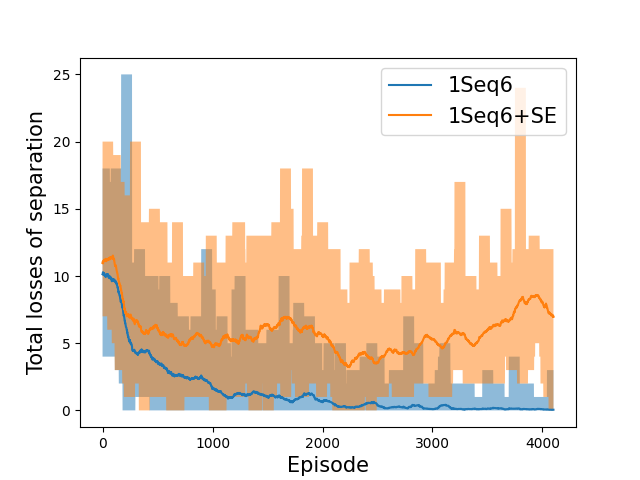}
     \end{subfigure}
     \hfill
     \begin{subfigure}[b]{0.40\textwidth}
         \centering
         \includegraphics[width=15pc]{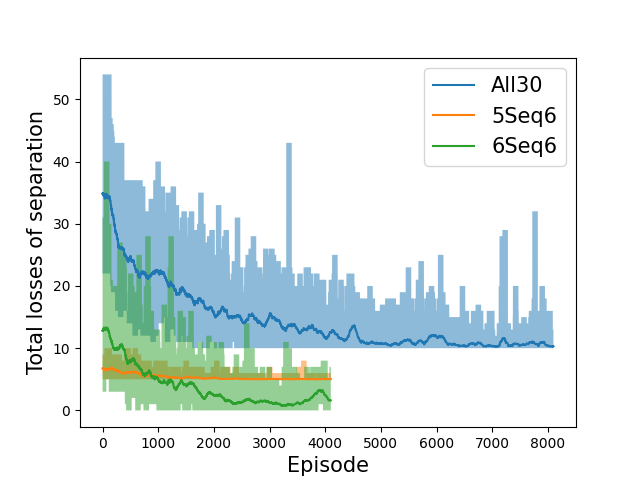}
     \end{subfigure} \\
     
        \caption{Learning curves showing the mean reward, number of resolution actions, alerts and losses per training episode. Left: for 1Seq6 using the proposed DGN with edges (1Seq6), and 1Seq6 using the DGN+SE variation (1Seq6+SE) and right: for All30, 5Seq6 and 6Seq6 using the proposed DGN with edges, only. }
\label{fig:curves}
\end{figure}

The learning curves for the models trained are provided in Figure \ref{fig:curves}: Curves provide the mean reward, number of resolution actions, alerts and losses. The mean reward is computed as the average reward for all agents, per scenario time step. Curves are provided (a) (left) for the 1Seq6 model learned by the proposed DGN with edges method, and the 1Seq6+SE  model; and (b) (right) For All30, 5Seq6 and 6Seq6 using the proposed DGN with edges method, and only while training the model with the final batch of scenarios. It must be noted that  5Seq6 and 6Seq6 curves end early as these models are are trained with 4000 episodes, while All30 is trained using 8000 episodes.
The curves on the left aim to show the benefits of adding the edges in DGN, while the curves on the right show the efficacy of the training patterns. Indeed, the 1Seq6+SE model does not manage to learn effectively compared to 1Seq6 (using edges), as it decreases agents' mean reward, increases significantly the resolution actions, causing the increase of alerts and losses of separation. 
Results on the right show how the efficacy of MSeq6 models increases, while increasing M: Indeed, 6Seq6 manages to increase agents' mean reward, reduce alerts, losses of separation, and resolution actions compared to All30, but with a slight increase of alerts and of  resolution actions compared to 5Seq6.

\begin{table}[]
    \centering
   \footnotesize
    \begin{tabular}{c!{\vrule width 1.1pt}c!{\vrule width 0.1mm}c!{\vrule width 1.1pt}c!{\vrule width 0.1mm}c!{\vrule width 1.1pt}c!{\vrule width 0.1mm}c}
        
        \multirow{2.3}{*}{\textbf{Model}} 
        & \multicolumn{2}{c!{\vrule width 1pt}}{\textbf{Conflicts resolved \%}}
        & \multicolumn{2}{c!{\vrule width 1pt}}{\textbf{Avg No. of Resolution Actions}}
        & \multicolumn{2}{c}{\textbf{Avg Additional NMs}}\\
        \cline{2-7}
        & {\ \ \ \ \ \ Train \ \ \ \ \ \ } & {Test} & {\ \ \ Train\ \ \ } & {Test} & {\ \ Train \ \ } & {Test} \\
        \ChangeRT{1.1pt}
        
        All30 & 96.08\% & 37.5\% & 2.36 & 3.17 & -1.07 & 0.13        \\
        1Seq6 & 100\% & 28.57\% & 5.0 & 3.5 & -0.4 & 0.97       \\
        5Seq6 & 80.81\% & 66.67\% & 5.07 & 3.00 & -1.33 & 3.25       \\
        6Seq6 & 80.65\% & 90.0\% & 4.0 & 2.5 & -0.01 & 0.35       \\
        \ChangeRT{1.1pt}
        
        \end{tabular}
    \normalsize
    \caption{Results of training and testing the  models.}
    \label{tab:results}
\end{table}

\section{Concluding remarks}
Aiming to contribute to operational integrity and automation adoption for  CD\&R in the ATM domain, this article proposes a solution based on the DGN graph convolutional reinforcement learning (RL)  method, enhanced with edges representing conflicts between flights. Experimental results show that the proposed  method, can provide high-quality solutions. Such solutions are also provided while the method works in interaction with the operational platform, in AoRs other than the Barcelona AoR in which it has been trained.
Also, the system has been designed for addressing  transparency, as required by ATCO experts, and sets the ground for further elaboration towards applying advanced AI methods in the ATC domain. 

Immediate plans include validating the system in simulated real-world settings with ATCO. Furthermore, enhancement of safety on automation by training the models in more scenarios so as to adapt efficiently to unseen test cases (e.g. by using adversarial models) is a necessity. In addition to this, enhancement of the reward function with additional terms, building and incorporating explainability functionality, always in cooperation with ATCO, to satisfy operational goals and promote trust and system acceptability, are within the future plans.
\\

\textbf{Acknowledgement} This work has been supported by the TAPAS H2020-SESAR-2019-2 Project (GA number 892358) Towards an Automated and exPlainable ATM System.

\bibliographystyle{plain} 
\bibliography{main}

\begin{thebibliography}{1}

\bibitem{brittain2019autonomous}
Marc Brittain and Peng Wei.
\newblock Autonomous air traffic controller: A deep multi-agent reinforcement
  learning approach.
\newblock In {\em IEEE ICTS}, 2019.

\bibitem{Dalmau2020ATC}
Ramon Dalmau and Eric Allard.
\newblock {Air Traffic Control Using Message Passing Neural Networks and
  Multi-Agent Reinforcement Learning}.
\newblock In {\em SID}, 2020.

\bibitem{EgorovATM21}
Maxim Egorov, Antony Evans, Scot Campbell, Sebastian Zanlongo, and Tyler Young.
\newblock Evaluation of utm strategic deconfliction through end-to-end
  simulation.
\newblock In {\em ATM Seminar}, 2021.

\bibitem{ghosh2020deep}
Supriyo Ghosh, Sean Laguna, Shiau~Hong Lim, Laura Wynter, and Hasan Poonawala.
\newblock A deep ensemble multi-agent reinforcement learning approach for air
  traffic control.
\newblock {\em arXiv:2004.01387}, 2020.

\bibitem{10.5555/3305381.3305512}
Justin Gilmer, Samuel~S. Schoenholz, Patrick~F. Riley, Oriol Vinyals, and
  George~E. Dahl.
\newblock Neural message passing for quantum chemistry.
\newblock In {\em ICML}, 2017.

\bibitem{jiang2020graph}
Jiechuan Jiang, Chen Dun, Tiejun Huang, and Zongqing Lu.
\newblock Graph convolutional reinforcement learning.
\newblock In {\em ICLR}, 2020.

\bibitem{li2019optimizing}
Sheng Li, Maxim Egorov, and Mykel Kochenderfer.
\newblock Optimizing collision avoidance in dense airspace using deep
  reinforcement learning.
\newblock In {\em ATM Seminar 2019}, 2019.

\bibitem{mythos}
Zachary~C. Lipton.
\newblock The mythos of model interpretability: In machine learning, the
  concept of interpretability is both important and slippery.
\newblock {\em Queue}, 16(3):31–57, Jun 2018.

\bibitem{pham2019machine}
Duc-Thinh Pham, Ngoc~Phu Tran, Sameer Alam, Vu~Duong, and Daniel Delahaye.
\newblock A machine learning approach for conflict resolution in dense traffic
  scenarios with uncertainties.
\newblock In {\em ATM Seminar 2019}, 2019.

\end{thebibliography}

\newpage
\appendix
\section{Appendix: Supplementary material}

\label{appendix}

\subsection{Details on scenarios for training and testing the models}

\FloatBarrier
\begin{table}[h]
    \centering
    \tiny
    \begin{tabular}{c|c|c|c}
       Scenario ID & Number of & Number of &  \\ 
         & flights &  conflicts/alerts/losses & Duration \\ \hline
       1564760140-LECBLVU & 32 & 9/3/1 & 1620 s  \\ 
       1564759880-LECBLVU & 31 & 10/4/1 & 1590 s \\
       1564741571-LECBVNI & 27 & 4/9/6 & 1980 s \\
       1564750070-LECBLVU & 38 & 2/0/0 & 1530 s \\
       1564751917-LECBP2R & 19 & 1/0/0 & 660 s \\
       1564736806-LECBVNI & 30 & 8/4/1 & 1920 s \\ 
       1564760500-LECBLVU & 31 & 4/2/1 & 1470 s \\
       1564759733-LECBBAS & 38 & 2/2/1 & 1710 s \\
       1564759684-LECBBAS & 36 & 2/2/1 & 1620 s \\
       1564752230-LECBP2R & 27 & 8/4/1 & 1620 s \\
       1564765001-LECBVNI & 31 & 7/0/0 & 1560 s \\
       1564731484-LECBP2R & 34 & 0/0/0 & 1320 s \\
       1564741469-LECBVNI & 29 & 5/9/6 & 1860 s \\
       1564773821-LECBCCC & 31 & 1/2/2 & 1590 s \\
       1564730565-LECBP2R & 34 & 0/0/0 & 1320 s \\
       1564751659-LECBCCC & 29 & 1/2/1 & 1530 s \\
       1564753001-LECBP2R & 27 & 8/4/1 & 1620 s \\
       1564759004-LECBBAS & 39 & 2/2/1 & 2760 s \\
       1564731484-LECBP2R & 34 & 0/0/0 & 1320 s \\
       1564753001-LECBP2R & 27 & 8/4/1 & 1620 s \\
       1564765000-LECBVNI & 31 & 7/0/0 & 1560 s \\
       1564745384-LECBP2R & 26 & 1/1/1 & 870 s \\
       1564777409-LECBP2R & 23 & 9/0/0 & 750 s \\
       1564745315-LECBP2R & 26 & 1/1/1 & 870 s \\
       1564734831-LECBBAS & 43 & 5/5/1 & 1800 s \\
       1564746345-LECBP2R & 43 & 1/2/1 & 1920 s \\
       1564772971-LECBCCC & 31 & 1/1/1 & 1530 s \\
       1564745170-LECBP2R & 39 & 1/2/1 & 1650 s \\
       1564740976-LECBVNI & 29 & 0/0/0 & 1800 s \\
       1564730330-LECBP2R & 32 & 0/0/0 & 1260 s \\
       1564731530-LECBP2R & 40 & 0/0/0 & 1740 s \\
       1564763981-LECBVNI & 54 & 13/4/0 & 2040 s \\
       1564765959-LECBBAS & 54 & 2/1/0 & 2070 s \\
       1564763894-LECBP2R & 28 & 4/2/0 & 1350 s \\
       1564736579-LECBGO3 & 34 & 11/0/0 & 1230 s \\
       1564741854-LECBVNI & 29 & 5/9/6 & 1860 s \\
       1564745446-LECBP2R & 26 & 1/1/1 & 870 s \\
       1564771680-LECBLVU & 29 & 0/1/1 & 1920 s \\
       1564735391-LECBBAS & 36 & 1/1/1 & 1560 s \\
       1564740452-LECBGO3 & 35 & 10/2/0 & 1230 s \\
       1564757225-LECBBAS & 31 & 1/1/0 & 1530 s \\
       1564756177-LECBLVU & 19 & 2/1/0 & 690 s \\\hline
    \end{tabular}
    \normalsize
    \caption{Scenarios for training/testing the models, in the order they have been used.}
    \label{tab:scenarios}
\end{table}
\FloatBarrier

\subsection{Technical details on edges and convolutions}

Given the agency matrix of any agent $i$ at time $t$, $C^t_i$ and the matrix $C^t$ of all agents (of size $(|Neigh(i,AoR,t)|+1)\times N$ and $N \times (|Neigh(i,AoR,t)|+1)\times N$, respectively, where $\textit{N}$ is the number of agents), the local observation $o^t_i$ of  agent $i$ is inserted into an MLP encoder, $\mathit{MLP}_{ObsEnc}$, and is projected in a higher dimensional space. The outcome is the feature vector $ho^t_i$ with length $\mathit{L}$. 

Additionally, the vector $e^t_{ij}$ of each agent $i$ is encoded by another MLP encoder $\mathit{MLP}_{EdgEnc}$, which outputs the feature vector $he^t_{ij}$ with length $\mathit{L}$, as well. It is significant to be mentioned that: The elements of vector $e^t_{ij}$ are sorted in the same way as in $C^t_i$, and the first element of vector $e^t_{ij}$, which is $e^t_{ii}$, consists of predefined values for each agent $i \in N$, to facilitate the calculations needed.  

Afterwards, for each agent $i$ and $j \in (Neigh(i,AoR,t) \cup \{i\})$ we concatenate the vectors $ho^t_j$ and $he^t_{ij}$ resulting in the combined vector $hc^t_{ij}$ with length $\mathit{2L}$:
$$
hc^t_{ij} = concat\big([ho^t_j, he^t_{ij}]\big)
$$
Considering $hc^t_{ij}$ as a row vector, we can stack them horizontally (for each agent $i$ separately) in order to obtain a matrix of vectors $hc^t_i$ of size $(|Neigh(i,AoR,t)|+1) \times 2L$. By merging all the agents' matrices, we can get the tensor $\mathit{HC}^t$ with shape $N \times (|Neigh(i,AoR,t)|+1) \times 2L$.
$$
hc^t_i = horizontalStack\Bigg(\Big[hc^t_{ij} \ \Big |\ \forall\ j \in \big(Neigh(i,AoR,t) \cup \{i\}\big)\Big]\Bigg)
$$
$$
\mathit{HC}^t = merge\Big(\big[hc^t_i\ \big|\ \forall\ i \in A\big]\Big)
$$
Technically speaking, a faster way to obtain $\mathit{HC}^t$ than the above analytic way, which exploits the full GPU capability, is to: (a) merge all the vectors $o^t_i$ of all agents into the vector $\mathit{O}^t$ and then to get $\mathit{HO}^t$ by passing it through $\mathit{MLP}_{ObsEnc}$, (b) then expand and tile (copy) its first dimension $\textit{N}$ times resulting in the tensor $\mathit{THO}^t$ with shape $N \times N \times L$, (c) stack horizontally all $e^t_{ij}$ for the neighbors of each agent, merge them into the tensor $\mathit{E}^t$ and get $\mathit{HE}^t$ (with shape $N \times (|Neigh(i,AoR,t)|+1) \times L$) by $\mathit{MLP}_{EdgEnc}$, (d) apply dot product on $C^t$ with $\mathit{HO}^t$ and concatenate the outcome with $\mathit{HE}^t$:
$$
 \mathit{O}^t = merge\Big(\big[o^t_i \ \big|\ \forall\ i \in A\big]\Big)
$$
$$
\mathit{HO}^t = \mathit{MLP}_{ObsEnc}(\mathit{O}^t)
$$
$$
\mathit{TH0}^t = tile\Big(expand\big(\mathit{HO}^t,\ axis=0\big),\ axis=0,\ times=N\Big)
$$
\begin{align*} 
\mathit{E}^t = merge\Bigg(\Bigg[horizontalStack\Big(\big[e^t_{ij}\ \big|\ \forall\ i \in A)\big]\Big)\ \Bigg |\\
\forall\ j \in \big(Neigh(i,AoR,t) \cup \{i\}\big)\Bigg]\Bigg)
\end{align*}
$$
\mathit{HE}^t = \mathit{MLP}_{EdgEnc}(\mathit{E}^t)
$$
$$
\mathit{HC}^t = concat\Big(\big[C^t \cdot (\mathit{HO}^t)^\intercal,\ \mathit{HE}^t\big]\Big)
$$

In the next steps, as  mentioned, the convolution kernel is the multi-head dot-product attention kernel, which can be computed as specified in section 4.2, for each agent $i$ and its neighbors:
\small
$$
a^l_{ij} = \frac{exp\Big(\big(W^l_{ii} \cdot (K^l_{ij})^\intercal\big) \times (1/\sqrt{d_{K^l_{ij}}})\Big)}{\sum_{j} exp\Big(\big(W^l_{ii} \cdot (K^l_{ij})^\intercal\big) \times (1/\sqrt{d_{K^l_{ij}}})\Big)}
$$
\normalsize
where $j$ varies in  $\big(Neigh(i,AoR,t) \cup \{i\}\big)$ $l$ denotes the corresponding convolution layer, $d_{K^l_{ij}}$ is equal to the size of each head in $K^l_{ij}$.  $W^l_{ii}$ and $K^l_{ij}$ are computed as follows:
$W^l_{ii} = f^l_W(hc^t_{ii})$
and
$K^l_{ij} = f^l_K(hc^t_{ij})$,
where $f^l_W$ and $f^l_K$ are one-layer MLPs with linear activation function. $W^l_{ii}$ is a matrix of size $m \times d_{W^l_{ij}}$ (note that $d_{W^l_{ij}}=d_{K^l_{ij}}$), where $m$ is the number of its heads. The same is true for  $K^l_{ij}$. Continuing the process, the attention values $a^l_{ij}$, which are represented as a vector with length $m$, are utilized to impose a weight on how significant is the neighbor-agent $j$ for agent $i$, which is reflected in the outcome of the convolution layer:
$$
\textstyle h^l_i = f^l_o\Bigg(\sum_{j \in \big(Neigh(i,AoR,t) \cup \{i\}\big)}a^l_{ij}  V^l_{ij}\Bigg)
$$
where $h^l_i$ is a vector with length $L$, $f^l_o$ is a one-layer MLP with ReLU activation function and $V^l_{ij}$, which is a matrix of size $m \times d_{V^l_{ij}}$ where $d_{V^l_{ij}}=d_{K^l_{ij}}$, can be obtained by:
$$V^l_{ij} = f^l_V(hc^t_{ij})$$
where $f^l_V$ is also a one-layer MLP (with linear activation function). By concatenating $h^l_i$ for all agents, we get the total output of the convolution layer $H^l = concat\big([h^l_i,\ \forall \ i \in A]\big)$, which is a matrix of size $N \times L$. 

Here again, we optimize operations as follows:
$$
A^l = softmax\Big(\big(W^l \cdot K^l\big) \times (1/\sqrt{d_{K^l_{ij}}})\Big)
$$    
$$
H^l = A^l \cdot V^l
$$

where $W^l$, $K^l$ and $V^l$ are computed as:
$$
W^l = f^l_W(RHCN^t)
$$
$$
K^l = f^l_K(RHC^t)
$$
$$
V^l = f^l_V(RHC^t)
$$
The tensor $RHC^t$ is a reshaped version of $HC^t$ with shape $N \times m \times (|Neigh(i,AoR,t)|+1) \times (2L/m)$, while $RHCN^t$ is a reshaped version of tensor $HCN^t$. The latter (with shape $N \times 2L$) is obtained by concatenating $\mathit{HO}^{t}$ and a slice of $HC^t$ (keeping only the first element of the second dimension): $HCN^t = concat\Big(\big[\mathit{HO}^t,\ slice(HC^t)\big]\Big)$. It should be clarified that the shape of tensor $W^l$ is $N \times m \times 1 \times d_{W^l_{ij}}$, while the shape of $K^l$ and $V^l$ (the same for both) is $N \times m \times (|Neigh(i,AoR,t)|+1) \times d_{W^l_{ij}}$. As a result, the shape of tensor $A^l$ is $N \times m \times 1 \times (|Neigh(i,AoR,t)|+1)$.

The operations of the second convolution layer are identical to the first, with $\mathit{FC}^l=concat\Big(\big[\mathit{C^t}\cdot (TH^l)^\intercal,\ \mathit{HE}^t\big]\Big)$ and $H^l$ as inputs (where $l=1$), and $H^2$ as output. The tensor $TH^l$ with shape $N \times N \times L$ is given by expanding and tiling the first dimension of the matrix $H^l$, $\textit{N}$ times: $TH^l = tile\big(expand(H^l,\ axis=0),\ axis=0,\ times=N\big)$. Note that the corresponding inputs of the first convolution layer were $\mathit{HO}^t$ and $\mathit{HC}^t$. The final step is to concatenate $\mathit{HC}^t$, $H^1$ and $H^2$ and feed them into the $Q$ network in order to obtain $Q-values$ of all actions:
$$
Q-values = Q-net\Big(concat\big([\mathit{HC}^t, H^1, H^2]\big)\Big)
$$

For the implementation of the baseline method DGN+SE where edges' features are incorporated into agents' observations,  the vector of the concatenated encoded observations features of the $j^{th}$ neighbor of agent $\mathit{i}$ is computed as follows: 

$hc^t_{ij}= \mathit{MLP}_{ObsEnc}\Bigg(concat\bigg(\Big[o^t_{j}, \big\{e^t_{jk} | \ k \in (Neigh(j,AoR,t)\big\}\Big]\bigg)\Bigg)$,

where $j \in (Neigh(i,AoR,t) \cup \{i\}$. In this case, the dot product of $W^l_{ii}$ with $K^l_{ij}$ combines $i$'s observations and the features of all of its neighbors, with those of its neighbor $j$. It must be noted that $j$'s edges features include all the information about $j$'s neighbors, except $i$.

\end{document}